\documentclass{elsarticle}
\usepackage{graphicx}
\usepackage{amsmath}
\begin{document}

\title{The Cubic Vortical Whitham Equation}

\author[su]{John D.~Carter\corref{cor1}}
\ead{carterj1@seattleu.edu}
\address[su]{Mathematics Department, Seattle University, USA}
\author[ber]{Henrik Kalisch}
\address[ber]{Department of Mathematics, University of Bergen, Norway}
\author[fra]{Christian Kharif}
\author[fra]{Malek Abid}
\address[fra]{Aix Marseille Universit\'e, CNRS, Centrale Marseille, IRPHE UMR 7342, F-13384, Marseille, France}
\cortext[cor1]{Corresponding author}

\begin{abstract}
    The cubic vortical Whitham equation is a model for wave motion on a vertically sheared current of constant vorticity in a shallow inviscid fluid.  It generalizes the classical Whitham equation by allowing constant vorticity and by adding a cubic nonlinear term.  The inclusion of this extra nonlinear term allows the equation to admit periodic, traveling-wave solutions with larger amplitude than the Whitham equation.  Increasing vorticity leads to solutions with larger amplitude as well.  The stability of these solutions is examined numerically.  All moderate- and large-amplitude solutions, regardless of wavelength, are found to be unstable.  A formula for a stability cutoff as a function of vorticity and wavelength for small-amplitude solutions is presented.  In the case with zero vorticity, small-amplitude solutions are unstable with respect to the modulational instability if $kh>1.252$, where $k$ is the wavenumber and $h$ is the mean fluid depth.  
\end{abstract}

\maketitle

\section{Introduction}
\label{Introduction}

The dimensional Korteweg-deVries equation~\cite{kdv} is a model for the evolution of small-amplitude long waves on a shallow inviscid fluid with a horizontal, impermeable bed.  It is given by
\begin{equation}
    \eta_t+\sqrt{gh}~\eta_x+\frac{1}{6}h^2\sqrt{gh}~\eta_{xxx}+\frac{3}{2h}\sqrt{gh}~\eta\eta_x=0,
    \label{KdVDimensional}
\end{equation}
where $\eta$ represents the surface displacement of the fluid, $g$ represents the acceleration due to gravity, and $h$ represents the mean depth of the fluid.   See Johnson~\cite{johnson} for a detailed derivation.  Applying the nondimensionalization
\begin{equation}
    \eta\rightarrow h\eta, \hspace*{1cm} x\rightarrow h x, \hspace*{1cm} t\rightarrow \sqrt{\frac{h}{g}}~t,
\label{nondimensionalization}
\end{equation}
gives the dimensionless Korteweg-deVries (KdV) equation,
\begin{equation}
    u_t+u_x+\frac{1}{6}u_{xxx}+\frac{3}{2}uu_x=0.
    \label{KdV}
\end{equation}
This is one of the most well-known nonlinear partial differential equations (PDEs).  It has been extensively studied both mathematically, see for example \cite{miles,Whithambook,AS,LannesBook}, and experimentally, see for example \cite{russell,zabusky,h,hs}).

The KdV equation only reproduces the linear phase velocity of the water-wave problem in the long-wave limit.  In order to address this deficiency, Whitham~\cite{Whitham,Whithambook} proposed the nonlocal equation that now bears his name
\begin{equation}
    u_t+K*u_x+\frac{3}{2}uu_x=0,
    \label{Whitham}
\end{equation}
where $K$ is the Fourier multiplier defined by
\begin{equation}
    K=\sqrt{\frac{\tanh(k)}{k}}.
    \label{WhithamK}
\end{equation}
The Whitham equation reproduces the unidirectional phase velocity of the full water-wave problem.  Here it is presented in dimensionless form.  Moldabayev~\cite{Moldabayev}, Trillo et al.~\cite{Trillo}, and Carter~\cite{WhithamComp} showed that the Whitham equation is a more accurate model for the evolution of waves of depression than the KdV and Serre equations.  Ehrnstr\"om \& Kalisch~\cite{EK} proved that the Whitham equation admits small-amplitude, periodic traveling-wave solutions and numerically computed periodic, travelling-wave solutions with a variety of amplitudes including those close to the highest wave.  Ehrnstr\"om \& Wahl\'en~\cite{WhithamCusp} proved Whitham's conjecture that (\ref{Whitham}) admits a highest, cusped traveling-wave solution.  Hur \& Johnson~\cite{HurJohnson2015} proved that small but finite amplitude periodic solutions of the Whitham equation are stable with respect to long wavelength perturbations when $kh<1.145$ and are unstable with respect to long wavelength perturbations when $kh>1.145$.  Here, $k$ is the wavenumber of the solution and $h$ is the fluid depth.  This result is qualitatively different than the Bottman \& Deconinck~\cite{BD} proof that {\emph{all}} traveling-wave solutions of KdV are stable.  However, it is qualitatively similar to the Benjamin \& Feir~\cite{BenF} result that establishes the stability cut-off as $kh=1.363$.  Sanford et al.~\cite{Sanford2014} numerically corroborated the Hur \& Johnson~\cite{HurJohnson2015} result and numerically showed that all large-amplitude periodic traveling-wave solutions of the Whitham equation are unstable regardless of their wavelength.

In recent years, generalizations of the Whitham equation have gained significant interest.  Aceves-S\'anchez et al.~\cite{ASMP}, Deconinck \& Trichtchenko~\cite{BernardOlga}, and Hur \& Pandey~\cite{HP} all proposed different generalizations of the Whitham equation that match the {\emph{bidirectional}} phase velocity of the water-wave problem.  Hur \& Johnson~\cite{Hur2015} studied a generalization of the Whitham equation that includes surface tension.  Dinvay et al.~\cite{Dinvay} showed that this capillary-Whitham equation gives a more accurate reproduction of the water-wave problem than the KdV and Kawahara (fifth-order KdV) equations.  Carter \& Rozman~\cite{STWhitham} presented results from a numerical study of solutions to the capillary-Whitham equation.  Herein, we focus on a generalization of the (unidirectional) Whitham equation that includes constant vorticity and cubic nonlinearity.

\begin{figure}
    \begin{center}
        \includegraphics[width=12cm]{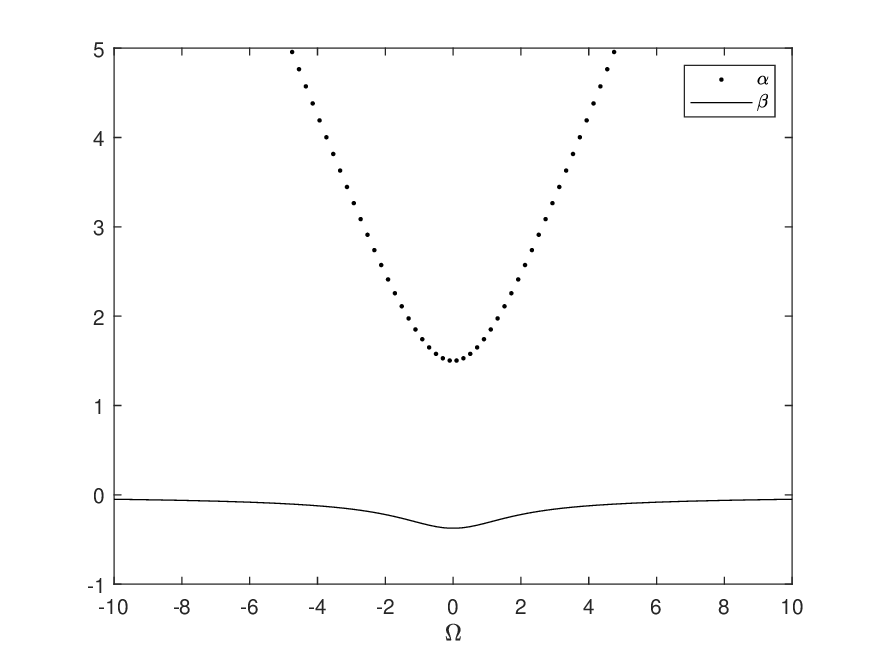}
        \caption{Plots of $\alpha$ and $\beta$ as functions of $\Omega$.}
        \label{AlphaBetaPlot}
    \end{center}
\end{figure}

Kharif et al.~\cite{Kharif2017} and Kharif \& Abid~\cite{Kharif2018} proposed an asymptotic model for fully nonlinear water waves propagating on a vertically sheared current of constant vorticity in shallow water that satisfies the unidirectional, linear dispersion relation. From this model they derived, within the framework of weakly nonlinear waves, a generalization of the Whitham equation which they named the vor-Whitham equation.  As a generalization of this, they proposed the following higher-order nonlinear model
\begin{equation}
    u_t+\mathcal{K}*u_x+\alpha uu_x+\beta u^2u_x=0,
    \label{CVW}
\end{equation}
where $\mathcal{K}$ is the Fourier multiplier defined by
\begin{equation}
    \mathcal{K}=-\frac{\Omega\tanh(k)}{2k}+\sqrt{\frac{\tanh(k)}{k}+\frac{\Omega^2\tanh^2(k)}{4k^2}},
    \label{CVWK}
\end{equation}
and the real constants $\alpha$ and $\beta$ are given by
\begin{subequations}
    \begin{equation}
        \alpha=\frac{\Omega^2+3}{\sqrt{\Omega^2+4}},
    \end{equation}
    \begin{equation}
        \beta=-\frac{6+\Omega^2}{2(4+\Omega^2)^{3/2}}.
    \end{equation}
\end{subequations}
As above, the implicit length and temporal scales are defined by $h$ and $\sqrt{h/g}$.  The parameter $\Omega$ represents the opposite sign of the constant vorticity.  Plots of $\alpha$ and $\beta$ versus $\Omega$ are included in Figure \ref{AlphaBetaPlot}.  We refer to equation (\ref{CVW}) as the cubic vortical Whitham or CV-Whitham equation.  Note that the Fourier multiplier $K$ in equation (\ref{WhithamK}) is the $\Omega=0$ version of the Fourier multiplier $\mathcal{K}$ given in equation (\ref{CVWK}).  The classical Whitham equation is obtained from (\ref{CVW}) and (\ref{CVWK}) by setting $\alpha=\frac{3}{2}$ and $\beta=\Omega=0$.

The CV-Whitham equation admits the following conserved quantities.
\begin{subequations}
    \begin{equation}
        \mathcal{Q}_1=\int_0^L u~dx,
        \label{CQ1}
    \end{equation}
    \begin{equation}
        \mathcal{Q}_2=\int_0^L u^2dx,
        \label{CQ2}
    \end{equation}
    \begin{equation}
        \mathcal{Q}_3=\frac{1}{2}\int_0^L\left(u\mathcal{K}*u+\frac{\alpha}{3}u^3+\frac{\beta}{6}u^4\right)dx.
        \label{CQ3}
    \end{equation}
\end{subequations}
The conserved quantity $\mathcal{Q}_3$ is the Hamiltonian.  These quantities are used as checks on the accuracy of our numerical results.

The remainder of this paper is organized as follows.  Section \ref{TWSolnSection} contains a description of the method we use to compute periodic, traveling-wave solutions to the CV-Whitham equation.  It also contains plots and descriptions of these solutions.  Section \ref{StabilitySection} contains the linear stability analysis along with numerical stability results.  Section \ref{Summary} contains a summary of the work.

\section{Traveling-wave solutions}
\label{TWSolnSection}
Consider periodic, traveling-wave solutions of the form
\begin{equation}
    u(x,t)=f(x-ct)=f(z),
    \label{TWAnsatz}
\end{equation}
where $c$ is a real constant and $f$ is a smooth, real-valued function of $z=x-ct$ with period $L$.  Substituting (\ref{TWAnsatz}) into (\ref{CVW}) and integrating with respect to $z$ once gives
\begin{equation}
    -cf+\mathcal{K}*f+\frac{\alpha}{2}f^2+\frac{\beta}{3}f^3=B,
    \label{TWEqn}
\end{equation}
where $B$ is the constant of integration.  Unlike the corresponding equation in the Whitham case, equation (\ref{TWEqn}) does not possess an invariant that allows the mean of the solution $f$ or the constant $B$ to be set to zero without loss of generality.  Nevertheless, we assume that the mean of the solution is zero because such solutions are most physically relevant.  Equation (\ref{TWEqn}) can be solved approximately by assuming $f$ has a Fourier expansion of the form
\begin{equation}
    f(z)=\sum_{k=-N}^{N} \hat{f}(k)\exp\left(\frac{2\pi ikz}{L}\right),
    \label{fForm}
\end{equation}
where $N$ is a large positive integer and the $\hat{f}$ are complex constants.  We assume that the solutions are real and even.  Therefore, all $\hat{f}$ are real valued and $\hat{f}(-k)=\hat{f}(k)$ for $k=1,\dots,N$.  Additionally, we impose $\hat{f}(0)=0$ so that the solutions have zero mean.  Substituting (\ref{fForm}) into (\ref{TWEqn}) gives the following system of equations
\begin{equation}
\begin{split}
    -c\hat{f}(k)&+\mathcal{K}\left(\frac{2\pi k}{L}\right)\hat{f}(k)+\frac{\alpha}{2}\left(\sum_{l=\max(-N,-N+k)}^{\min(N,N+k)}\hat{f}(k-l)\hat{f}(l)\right)\\ &+\frac{\beta}{3}\left(\sum_{j=\max(-2N,-N+k)}^{\min(2N,N+k)}\sum_{l=\max(-N,-N+j)}^{\min(N,N+j)}\hat{f}(k-j)\hat{f}(j-l)\hat{f}(l)\right)=0,\hspace*{0.5cm}\text{for $k=1,\dots,N$}.
\end{split}
\label{NewtonSystem}
\end{equation}
The case $k=0$ is not included in this system because it is used to determine the integration constant $B$ and is otherwise unimportant.  Equation (\ref{NewtonSystem}) is a system of $N$ equations for the $N+1$ unknowns $\hat{f}(k)$ for $k=1,\dots,N$ and $c$.  A solution to this system is found using linear theory and Newton's method.  Given values for $L$ and $\Omega$, choose a value of $c$ slightly larger than $c=\mathcal{K}\left(\frac{2\pi}{L}\right)$ and then solve (\ref{NewtonSystem}) using Newton's method with $\hat{f}(1)=0.1$ and $\hat{f}(k)=0$ for $k=2,\dots,N$ as the initial estimate for the solution.  This gives the Fourier coefficients of a small-amplitude, $L$-periodic solution to the CV-Whitham equation.  Fourier coefficients of solutions with larger amplitudes are found by incrementing $c$ by a small amount and solving (\ref{NewtonSystem}) via Newton's method using the $\hat{f}$ values from the previous solution as the initial estimate.

\begin{figure}
    \begin{center}
        \includegraphics[width=16cm]{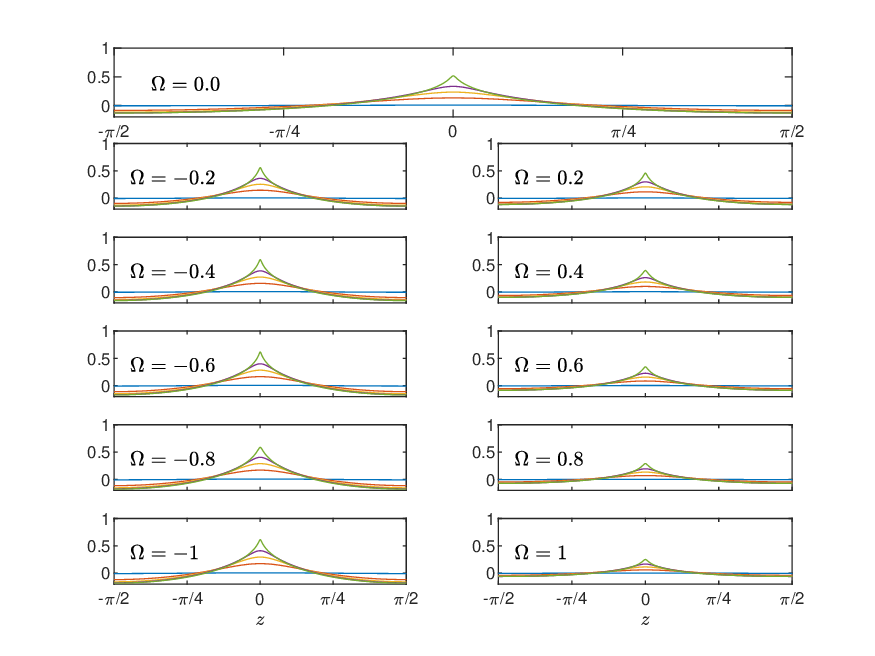}
        \caption{Traveling-wave solutions to the CV-Whitham equation with period $L=\pi$ for $\Omega=0,\pm0.2, \pm0.4, \pm0.6, \pm0.8, \pm1.0$.}
        \label{s314SolnPlot}
    \end{center}
\end{figure}

\begin{figure}
    \begin{center}
        \includegraphics[width=16cm]{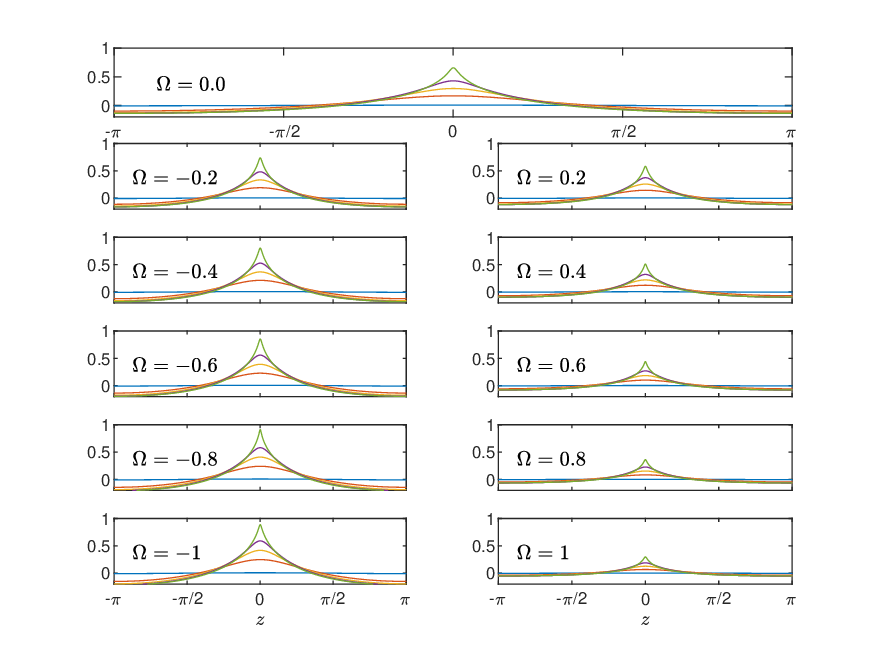}
        \caption{Traveling-wave solutions to the CV-Whitham equation with period $L=2\pi$ for $\Omega=0,\pm0.2, \pm0.4, \pm0.6, \pm0.8, \pm1.0$.}
        \label{s628SolnPlot}
    \end{center}
\end{figure}

This procedure can be used to find families of solutions to the CV-Whitham equation for any $\Omega$ and (positive) $L$.  Representative plots of solutions are included in Figures \ref{s314SolnPlot} and \ref{s628SolnPlot}.  Figure \ref{s314SolnPlot} contains plots of solutions to the CV-Whitham equation when $L=\pi$ and $\Omega=0.0, \pm0.2, \pm0.4, \pm0.6, \pm0.8, \pm1.0$ and Figure \ref{s628SolnPlot} contains plots of solutions when $L=2\pi$ for the same values of $\Omega$.  These plots (and others omitted for conciseness) show that the families of solutions are qualitatively similar regardless of the values of $L$ and $\Omega$.  Small-amplitude solutions appear very trigonometric and solution steepness increases with wave height.  For each $L$ and $\Omega$, there appears to be a solution with maximal wave height.  This solution is very steep at its peak and may be cusped.  However since our numerics are based on Fourier series, it is difficult to accurately resolve the precise shape of the solution with maximal wave height.  Figures \ref{s314SolnPlot} and \ref{s628SolnPlot} also show that for a given period, as $\Omega$ increases, the wave height of the solution with maximal wave height decreases in magnitude.


\begin{figure}
    \begin{center}
        \includegraphics[width=16cm]{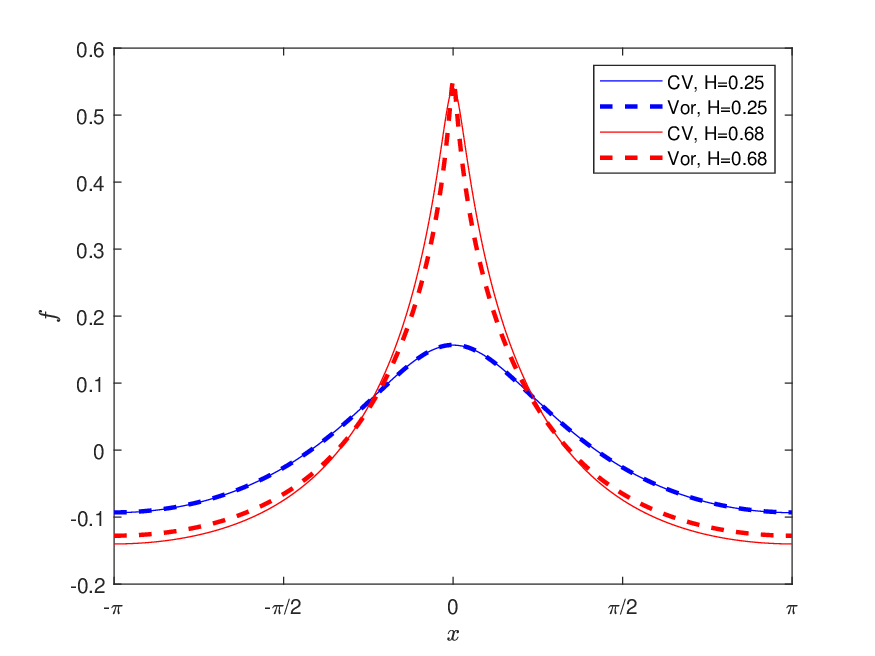}
        \caption{Traveling-wave solutions to the CV-Whitham equation (solid curves) and the Vor-Whitham equation (dashed curves) with $L=2\pi$, $\Omega=0.0$, and wave height $H=0.25, 0.68$.}
        \label{CVversusVor}
    \end{center}
\end{figure}

The CV-Whitham equation examined herein is similar to the vor-Whitham equation examined in Kharif et al.~\cite{vorW}.  The only difference is the addition of the cubic nonlinear term in (\ref{CVW}).  In essence, the CV-Whitham equation is ``more nonlinear'' than the vor-Whitham equation.  For given values of $L$ and $\Omega$, the difference between periodic traveling-wave solutions of the CV-Whitham and vor-Whitham equations is not large, see Figure \ref{CVversusVor}.  In this plot, the CV-Whitham solutions have speeds $c=0.8943$ (for $H=0.25$) and $c=0.9768$ (for $H=0.68$).  The vor-Whitham solutions have speeds $c=0.8960$ (for $H=0.25$) and $c=0.9757$ (for $H=0.68$).  Note that the difference between the two large-amplitude solutions in the plot is significant while the difference between the two small-amplitude solutions is not visible using this vertical scale.  The difference between the solutions to the CV-Whitham and vor-Whitham equations decreases as $\Omega$ increases in magnitude.  Note that the magnitude of $\beta$, the coefficient of the cubic nonlinear term, is maximized when $\Omega=0$.  For given values of $L$ and $\Omega$, the solution of maximal height of the CV-Whitham equation appears to have a larger wave height than the wave of maximal height of the vor-Whitham equation.  This means that the added nonlinearity allows solutions of larger amplitude.  

\begin{figure}
    \begin{center}
        \includegraphics[width=16cm]{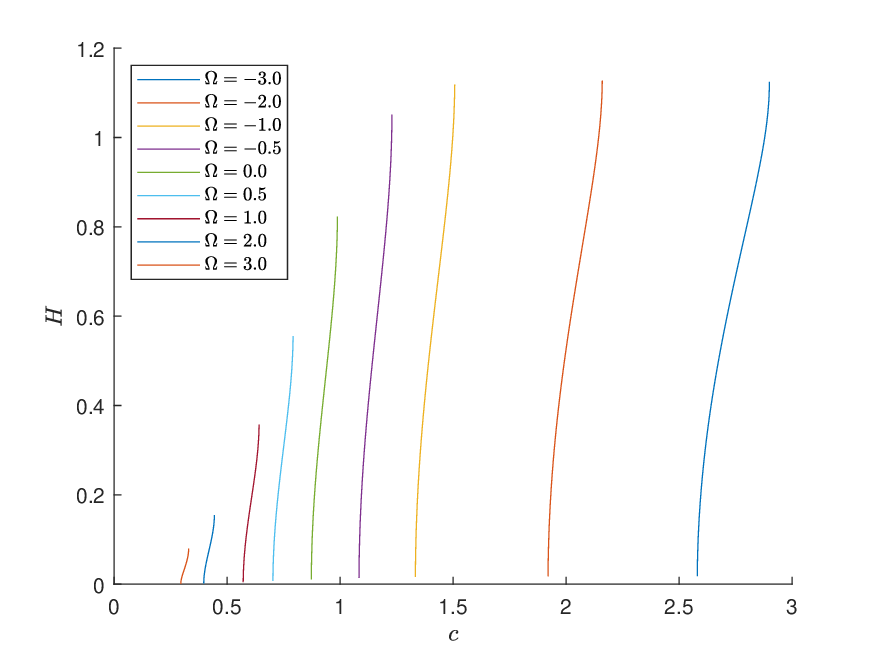}
        \caption{Plot of wave height, $H$, versus wave speed, $c$, for solutions to the CV-Whitham equation with period $L=2\pi$ and $\Omega=0.0, \pm0.5, \pm1.0, \pm2.0, \pm3.0$.  The curves from left to right are ordered by decreasing values of $\Omega$.}
        \label{s628Bifurcation}
    \end{center}
\end{figure}

\begin{figure}
    \begin{center}
        \includegraphics[width=16cm]{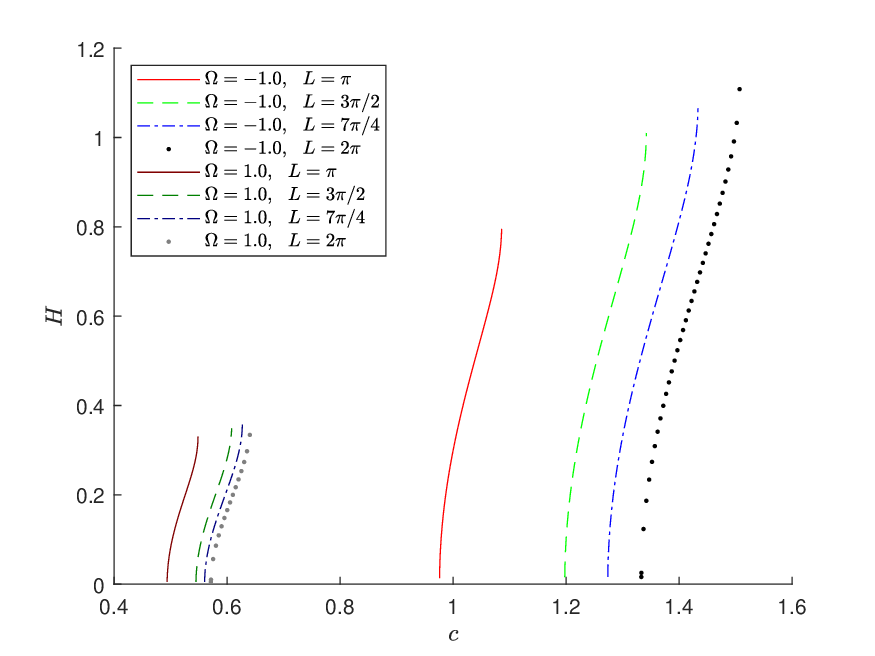}
        \caption{Plots of wave height, $H$, versus wave speed, $c$, for solutions to the CV-Whitham equation with $\Omega=\pm1.0$ and $L=\pi, \frac{3\pi}{2}, \frac{7\pi}{2}, 2\pi$.  The four curves on the left correspond to $\Omega=1.0$ and the four on the right correspond to $\Omega=-1.0$.}
        \label{OtherBifurcation}
    \end{center}
\end{figure}

Figures \ref{s628Bifurcation} and \ref{OtherBifurcation} include plots of the bifurcation curves showing wave height, $H$, as a function of wave speed, $c$.  These plots show that for given values of $L$ and $\Omega$, as wave speed increases, so does wave height.  For a fixed value of $L$, the maximal wave height increases as $\Omega$ decreases.  However, as $\Omega$ decreases below $-1.0$, the increase in maximal wave height is small.  For a fixed value of $\Omega$, the maximal wave height increases as the period increases.  The difference between the bifurcation plots of the CV- and vor-Whitham equations is small.  There is not a simple relationship between the phase speeds of the two equations, though the difference increases as $\Omega$ decreases.

\begin{figure}
    \begin{center}
        \includegraphics[width=16cm]{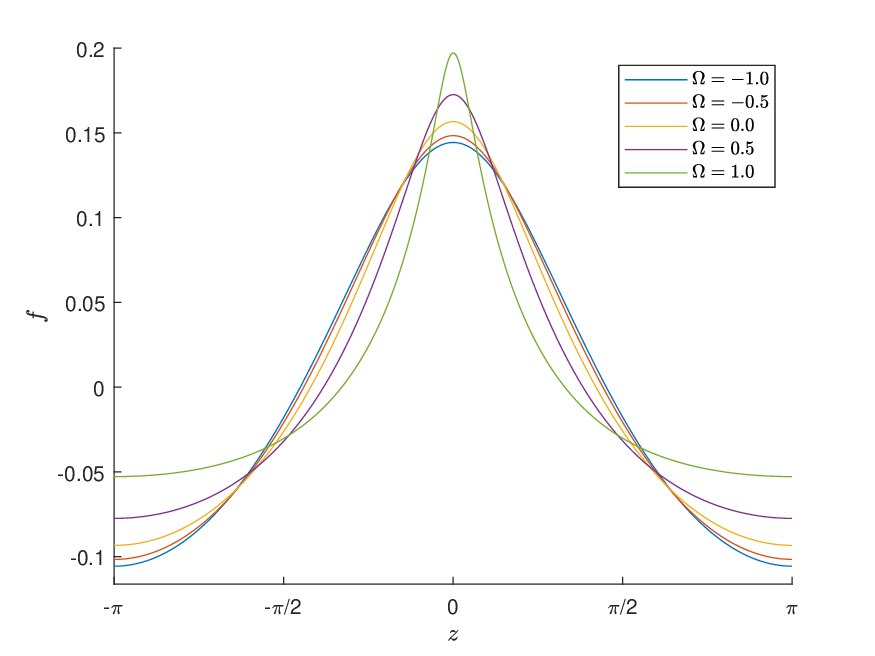}
        \caption{Plots of traveling-wave solutions to the CV-Whitham equation with period, $L=2\pi$, and wave height, $H=0.25$, for $\Omega=0.0, \pm0.5, \pm1.0$.}
        \label{SameHeightPlot}
    \end{center}
\end{figure}

Figure \ref{SameHeightPlot} includes plots of solutions with period $L=2\pi$ and wave height of $H=0.25$ for five different values of $\Omega$.  The plots show that as the value of $\Omega$ increases, the wave steepness also increases.  The same trend is observed in the vor-Whitham equation, see Kharif et al.~\cite{vorW}.

\section{Stability}
\label{StabilitySection}

In order to study the stability of traveling-wave solutions to the CV-Whitham equation, we use the Fourier-Floquet-Hill method of Deconinck \& Kutz~\cite{DK}.  In order to do this, first make the change of variables defined by
\begin{equation}
    z=x-ct,\hspace*{2cm}\tau=t.
    \label{COV}
\end{equation}
This converts the CV-Whitham equation to
\begin{equation}
    u_{\tau}-cu_z+\mathcal{K}*u_z+\alpha uu_z+\beta u^2u_z=0,
    \label{MovingCVW}
\end{equation}
and converts the traveling-wave solutions of (\ref{CVW}) into stationary solutions of (\ref{MovingCVW}).  In order to study the stability of these solutions, consider perturbed solutions of the form
\begin{equation}
    u_{pert}(z,\tau)=u(z)+\epsilon v(z,\tau)+\mathcal{O}(\epsilon^2),
    \label{PertForm}
\end{equation}
where $u$ is one of the solutions introduced in the previous section and $\epsilon v$ is the leading-order part of the perturbation.  Substituting (\ref{PertForm}) into (\ref{MovingCVW}) and linearizing gives
\begin{equation}
    v_\tau-cv_z+\mathcal{K}*v_z+(\alpha u^{\prime}+2\beta uu^{\prime})v+(\alpha u+\beta u^2)v_z=0,
    \label{LinearProb}
\end{equation}
where ``prime'' means derivative with respect to $z$.  Since this nonlocal PDE is constant coefficient in $\tau$, without loss of generality assume
\begin{equation}
    v(z,\tau)=V(z)\mbox{e}^{\lambda\tau}+c.c.,
    \label{vForm}
\end{equation}
where $V$ is a complex-valued function, $\lambda$ is a complex constant, and $c.c.$ stands for complex conjugate.  Substituting (\ref{vForm}) into (\ref{LinearProb}) and rearranging gives
\begin{equation}
    \mathcal{L}V=\lambda V,
    \label{LinearProb2}
\end{equation}
where $\mathcal{L}$ is the nonlocal differential operator defined by
\begin{equation}
    \mathcal{L}=c\partial_z-\mathcal{K}*\partial_z-(\alpha u^{\prime}+2\beta uu^{\prime})-(\alpha u+\beta u^2)\partial_z.
    \label{L}
\end{equation}
Equation (\ref{LinearProb2}) is a differential eigenvalue problem with eigenvalue $\lambda$ and eigenfunction $V$.  Floquet's Theorem~\cite{DK} states that all bounded solutions to this equation have the form
\begin{equation}
    V(z)=\mbox{e}^{i\mu z}\sum_{k=-N}^{N}\hat{V}(k)\mbox{e}^{2\pi ikz/L}
    \label{VForm}
\end{equation}
where the $\hat{V}(k)$ are complex numbers and $\mu\in[-\frac{\pi}{L},\frac{\pi}{L}]$ is a real constant known as the Floquet parameter.  If any $\lambda$ has a positive real part, then the corresponding perturbation grows exponentially in $\tau$ and the solution, $u(z)$, is said to be unstable.  If all $\lambda$ are purely imaginary, then the solution is said to be spectrally stable because there are no perturbations that grow exponentially.

Substituting (\ref{VForm}) into (\ref{LinearProb2}) and truncating gives
\begin{equation}
    \hat{\mathcal{L}}\hat{\mathcal{V}}=\lambda\hat{\mathcal{V}},
    \label{EvalProb}
\end{equation}
where $\hat{\mathcal{V}}=(\hat{V}(-N),\hat{V}(-N+1),\dots,\hat{V}(N))^T$ and the elements of the matrix $\hat{\mathcal{L}}$ are given by
\begin{equation}
    \hat{\mathcal{L}}_{mn}=  \begin{cases} 
        i\left(\mu+\frac{2\pi m}{L}\right)\left(c-\mathcal{K}\left(\mu+\frac{2\pi m}{L}\right)\right), & m=n, \\
        i\left(\mu+\frac{2\pi m}{L}\right)\left(-\alpha\hat{u}(n-m)-\beta\hat{w}(n-m)\right), & m\ne n,
     \end{cases}
    \label{LHat}
\end{equation}
where $\hat{w}$ are the Fourier coefficients of $u^2$.  Given values of $L$ and $\Omega$ along with a solution, we choose an equally-spaced sampling of $\mu$ values and find the corresponding eigenvalues and eigenvectors using a standard eigensolver (``eigen'' in the Julia programming language).  

\begin{figure}
    \begin{center}
        \includegraphics[width=16cm]{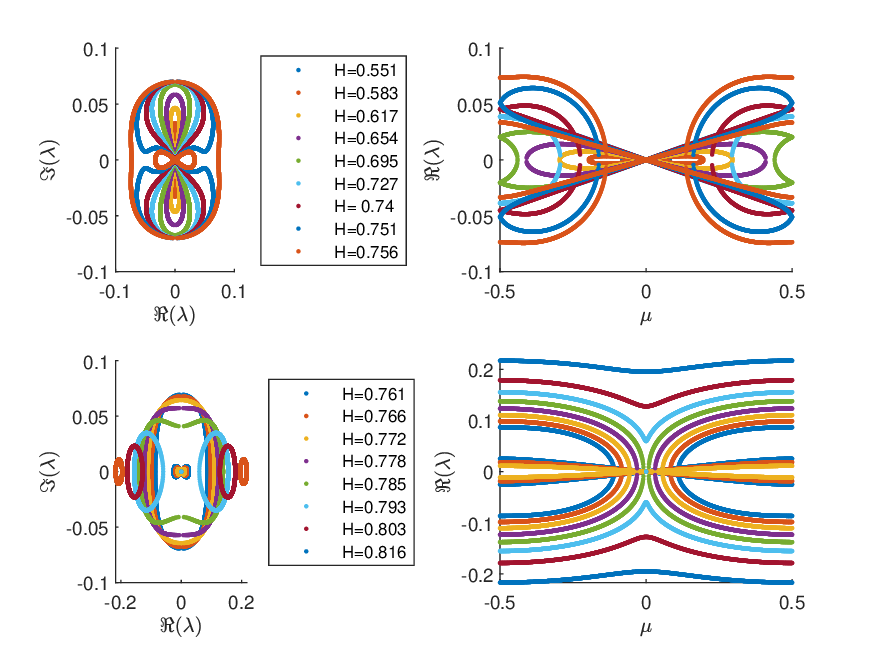}
        \caption{Plots of the stability spectra (left) and growth rate versus $\mu$ (right) for a number of solutions to the CV-Whitham equation with $L=2\pi$ and $\Omega=0.0$.  The second row of plots correspond to solutions with wave heights that are larger than those in the first row.}
        \label{s628Om0StabPlot}
    \end{center}
\end{figure}

Figure \ref{s628Om0StabPlot} contains plots of the stability spectra (in the left column) and the growth rate versus the Floquet parameter (in the right column) for solutions with $L=2\pi$, $\Omega=0.0$, and a number of different wave heights.  The plots corresponding to solutions of the CV-Whitham equation corresponding to other values of $L$ and $\Omega$ are qualitatively similar to those shown in Figure \ref{s628Om0StabPlot}.  In the $L=2\pi$ and $\Omega=0.0$  case, all solutions with wave height less than $H=0.5426$ or speed less than $c=0.9524$ are spectrally stable.  Solutions with larger wave heights or speeds are unstable.  Figure 3 of McLean's~\cite{McLean} classic work shows that the wave height cutoff for the full water-wave problem is between 0.4 and 0.58.  The cutoff value for the CV-Whitham equation is within this range.  (A more precise comparison cannot be made because \cite{McLean} includes a plot instead of a table of values.)  We note that our numerical exploration of $(L, \Omega)$-space showed that all large-amplitude solutions of the CV-Whitham equation are unstable regardless of wavelength.  

The plots in Figure \ref{s628Om0StabPlot} show that as the wave height or speed of the solution increases, so does the maximum instability growth rate.  This means that solutions with larger wave height are more unstable than solutions with smaller wave height.  The plots also show that if a solution with ``moderate'' wave height is unstable, then there is a ``figure eight'' centered at the origin of the spectrum.  For solutions with ``large'' wave height, there is a ``figure infinity'' centered at the origin of the spectrum surrounded by an oval.  As the wave height continues to increase, the size of the figure infinity decreases until it appears to collapse to the origin (i.e.~becomes too small to resolve using double precision computations).  Simultaneously, the outer oval pinches off along the vertical axis and becomes two smaller ovals.  For solutions with very large wave height, there are two oval-like shapes centered on the real axis.  As wave height increases, these ovals decrease in size while their centers trend away from the origin while remaining centered on the real axis.  It is unclear if the centers of these ovals head to $\Re(\lambda)=\pm\infty$ or to finite values as wave height increases.  

The plots of the growth rate versus the Floquet parameter show that solutions with large enough wave height are unstable with respect to a perturbation with $\mu=0.0$.  This means that such solutions are unstable with respect to perturbations that have the same wavelength as the solution.  Such instabilities are known as superharmonic instabilities.  These instabilities are quite different than the modulational (long-wave) instability that has been proven to exist in the small-amplitude limit of the Whitham equation, see~\cite{HurJohnson2015}.  If $\Omega=0.0$ and $L=2\pi$, solutions with $H\ge0.78976$ or $c\ge0.98735$ are unstable with respect to perturbations with period $2\pi$.  We were unable to find a simple relationship between $(\Omega,L)$ and the $(H,c)$ value of this cutoff.

\begin{figure}
    \begin{center}
        \includegraphics[width=16cm]{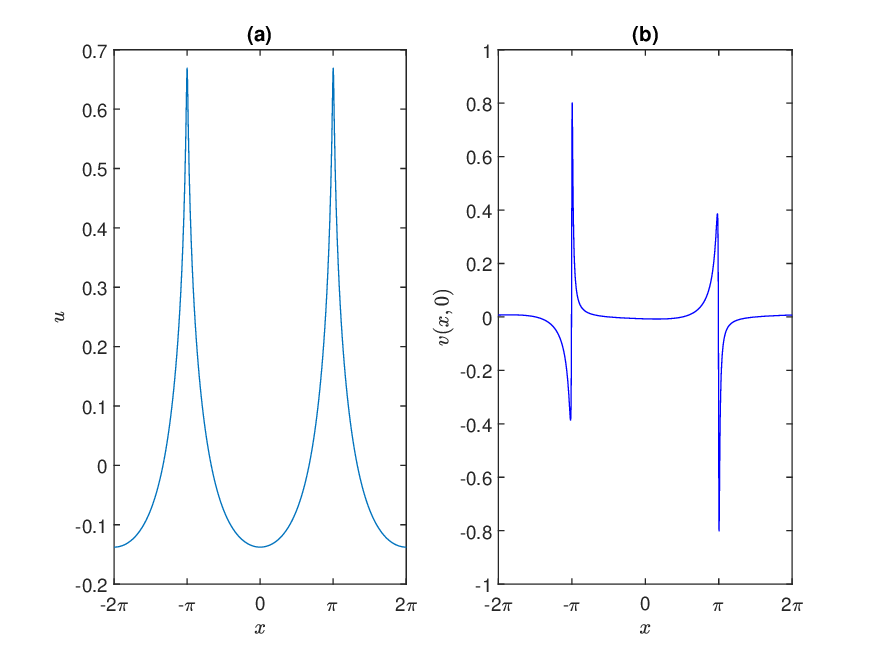}
        \caption{Plots of (a) two periods of a solution to the CV-Whitham equation with $L=2\pi$, $\Omega=0.0$, $H=0.8065$, and $c=0.9880$ and (b) the corresponding most unstable perturbation, $v(x,0)$, when $\mu=0.5$.}
        \label{StabCheck}
    \end{center}
\end{figure}

As an independent check on the results shown in Figure \ref{s628Om0StabPlot}, we selected the traveling-wave solution to the CV-Whitham equation with $\Omega=0.0$, $H=0.8065$, and $c=0.9880$, see Figure \ref{StabCheck}(a), and an unstable perturbation with $\mu=0.5$ and $\lambda=0.18905$, see Figure \ref{StabCheck}(b).  The unstable mode is localized near the peak of the traveling-wave solution.  This is the characteristic feature of the crest instability, also known as the Tanaka instability, see Tanaka {\emph{et al.}}~\cite{Tanaka}.  Two periods of the solution are shown because the period of the perturbation is twice that of the solution when $\mu=0.5$.  Then we used
\begin{equation}
    u_{pert}(x,0)=u(x)+10^{-7}v(x,0)
    \label{PertSoln}
\end{equation}
as an initial condition in our CV-Whitham solver while monitoring the growth of the perturbation.  The perturbation grew exponentially with rate of approximately $r=0.18905$ until its amplitude was large enough that nonlinear effects began to play a role.  This result corroborates the value of one point in Figure \ref{s628Om0StabPlot}.  

\begin{figure}
    \begin{center}
        \includegraphics[width=16cm]{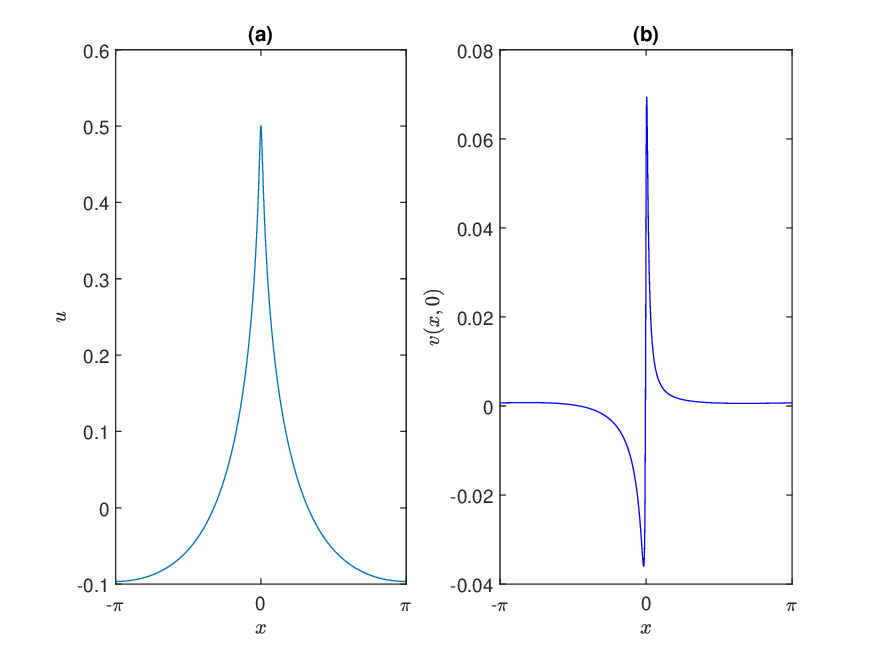}
        \caption{Plots of (a) one period of the solution to the CV-Whitham equation with $L=2\pi$, $\Omega=0.40$, $H=0.5969$, and $c=0.827233$ and (b) the corresponding unstable perturbation, $v(x,0)$, when $\mu=0.0$.}
        \label{StabCheckmu0}
    \end{center}
\end{figure}

For comparative purposes, we let $\Omega=0.4$ and selected the traveling-wave solution to the CV-Whitham equation with $H=0.5969$ and $c=0.8272$, see Figure \ref{StabCheckmu0}(a).  The stability analysis described in Section \ref{StabilitySection} with $\mu=0.0$, i.e.~only considering perturbations with the same as the unperturbed solution, led to a single unstable perturbation with $\lambda=0.1540$, see Figure \ref{StabCheckmu0}(b).  This unstable mode also exhibits similarity to the crest/Tanaka instability.  The perturbations shown in Figures \ref{StabCheck}(b) and \ref{StabCheckmu0}(b) are similar, but the perturbation in the $\mu=0.0$ case effects all wave peaks in the exactly same manner.  Using equation (\ref{PertSoln}) as an initial condition in this case led to the perturbation growing exponentially with a rate of approximately $r=0.1530$.  Similar checks were made for a handful of other points including some with complex $\lambda$ values.  The agreement between $\lambda$ (growth rate computed via the stability analysis) and $r$ (growth rate measured by computing the time evolution of the perturbed solution) was typically in the range of three to five decimal places.

\begin{figure}
    \begin{center}
        \includegraphics[width=16cm]{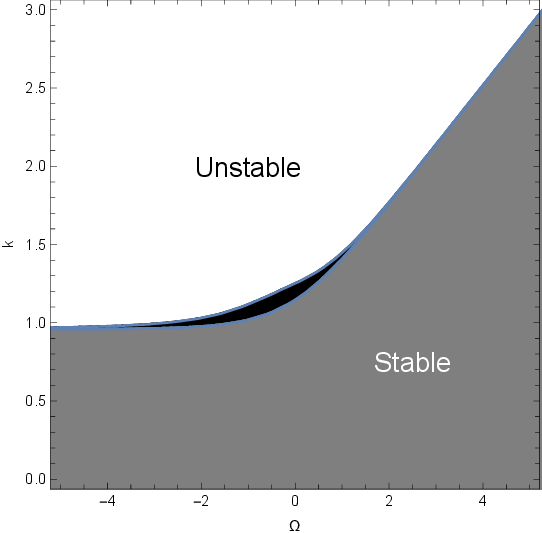}
        \caption{Plots of the stability region in the $(\Omega,k)$-plane for the CV-Whitham and vor-Whitham equations.  The black region corresponds to $(\Omega,k)$ values for which small-amplitude, periodic, traveling-wave solutions to the CV-Whitham equation are stable with respect to the modulational instability.  The gray region shows where small-amplitude, periodic, traveling-wave solutions to both the CV-Whitham and vor-Whitham equations are stable.}
        \label{CutoffPlot}
    \end{center}
\end{figure}

Benjamin~\cite{B} showed that small-amplitude, periodic traveling waves are unstable with respect to long-wavelength perturbations (i.e.~the modulational instability) if the wave number is greater than $k=1.363$ and are stable with respect to long-wavelength perturbations otherwise.  Hur \& Johnson~\cite{HurJohnson2015} proved that this cutoff occurs at $k=1.146$ for the Whitham equation.  Hur \& Johnson~\cite{Hur2015} derived a formula for the boundary between small-amplitude solutions of the vor-Whitham equation that are stable and unstable with respect to the modulational instability.  This region is represented by the gray area in Figure \ref{CutoffPlot}.  Binswanger et al.~\cite{Binswanger} derived a general formula that can be used for a wide variety of Whitham-type equations.  This formula states that if
\begin{equation}
    n(0,k)\Psi^{\prime\prime}(k)<0,
    \label{BinswangerFormula}
\end{equation}
where
\begin{subequations}
    \begin{equation}
        n(0,k)=-\left(\frac{(g^{\prime\prime}(0))^2}{\Psi^{\prime}(k)}+\frac{k(g^{\prime\prime}(0))^2}{2\Psi(k)-\Psi(2k)}+\frac{1}{2}g^{\prime\prime\prime}(0)\right)k,
    \end{equation}
    \begin{equation}
        \Psi(k)=k\mathcal{K}-\left(-\frac{\Omega}{2}+\sqrt{1+\frac{\Omega^2}{4}}\right)k,
    \end{equation}
    \begin{equation}
        g(y)=\frac{\alpha}{2}y^2+\frac{\beta}{3}y^3,
    \end{equation}
    \label{BinswangerFormula2}
\end{subequations}
then small-amplitude solutions of the CV-Whitham equation are stable with respect to the modulational instability.  Plots of the stability (with respect to the modulational instability) regions in the $(\Omega,k)$-planes for the CV-Whitham and vor-Whitham equations are included in Figure \ref{CutoffPlot}.  The plots show that the stability region for the CV-Whitham equation is slightly larger than that of the vor-Whitham equation.  Thus, the addition of the cubic nonlinear term provides a small stabilizing effect.  We numerically corroborated these results by computing the stability of solutions near the cutoff boundaries.  

Equations (\ref{BinswangerFormula}) and (\ref{BinswangerFormula2}) establish that when $\Omega=0$ in the CV-Whitham equation, the stability cutoff occurs at $k=1.252$.  This means that the CV-Whitham cutoff is closer to the ``exact'' cutoff of $k=1.363$ than the $k=1.146$ cutoff obtained from the Whitham equation.  Recall that the CV-Whitham equation is obtained from a Taylor series expansion of equation (20) of Kharif \& Abid~\cite{Kharif2018}.  Unfortunately, a cutoff for that equation cannot be determined using the Binswanger et al.~\cite{Binswanger} methodology because it does not have the required form.

\begin{figure}
    \begin{center}
        \includegraphics[width=16cm]{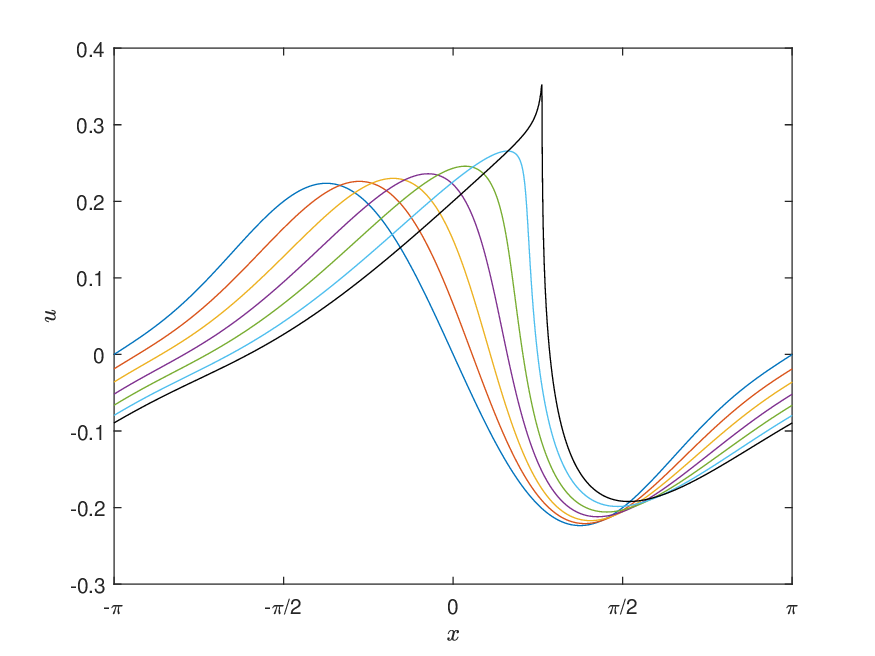}
        \caption{Plots of $u(x,t)$ versus $x$ for seven $t$ values for the initial condition given in equation (\ref{PeakingIC}) with $a=0.20$.}
        \label{PeakingPlot}
    \end{center}
\end{figure}

Finally, we examine peaking by considering an initial condition of the form
\begin{equation}
    u(x,0)=a\cos\left(x+\frac{\pi}{2}\right)+\frac{3-\tanh^2(1)}{4\tanh^3(1)}a^2\cos\left(2x+\frac{\pi}{2}\right),
    \label{PeakingIC}
\end{equation}
where $a=0.20$ in the CV-Whitham equation with $\Omega=1.0$.  This initial condition corresponds to a second-order Stokes wave on water with a (dimensionless) depth of 1, see for example Johnson~\cite{johnson}.  Figure \ref{PeakingPlot} shows the evolution of this initial condition for a variety of $t$ values.  Peaking occurs shortly after $t=1.95$.  As the peaking time is approached, both the asymmetry and wave height of the solution increase.  For $t$ values close to peaking, the numerics show the incipient formation of a vertical jet at the crest of the front of the wave.  Increasing the value of $a$ in equation (\ref{PeakingIC}) leads to earlier peaking times.  Decreasing the value of $a$ leads to either later peaking times or no peaking at all.  Increasing the value of $\Omega$ leads to earlier peaking times, while decreasing $\Omega$ leads to either later peaking times or no peaking at all.  

\section{Summary}
\label{Summary}

We have examined periodic traveling-wave solutions of the cubic vortical Whitham equation.  We showed numerically that the addition of a cubic nonlinear term allows the solutions to achieve larger wave heights.  We showed that increasing vorticity (decreasing $\Omega$) also leads to solutions with larger amplitudes.  We showed that all moderate-amplitude solutions are unstable and that some large-amplitude solutions are unstable with respect to perturbations with the same wavelength as the solution.  We showed that small-amplitude solutions of the CV-Whitham equation are stable with respect to the modulational instability for larger ranges of vorticity and wavelength than in the vortical Whitham equation.  Finally, we showed that the CV-Whitham equation more accurately reproduces the $kh=1.363$ modulational instability cutoff than does the Whitham equation.

\section{Acknowledgements}

We thank Bernard Deconinck for helpful discussions.  This material is based upon work supported by the National Science Foundation under Grant No. DMS-1716120 (JDC).  Malek Abid and Christian Kharif acknowledge the funding by the Excellence Initiative of Aix-Marseille University—A*Midex, a French “Investissements d’Avenir programme” AMX-19-IET-010.  Additionally, this work was supported by a CIRM Research in Pairs Grant that allowed JDC and HK to visit CK and MA in Luminy, France.


\begin{thebibliography}{10}

    \bibitem{AS}
    M.J. Ablowitz and H.~Segur.
    \newblock {\em Solitons and the Inverse Scattering Transform}.
    \newblock SIAM, Philadelphia, 1981.
    
    \bibitem{ASMP}
    P.~Aceves-S\'anchez, A.A. Minzoni, and P.~Panayotaros.
    \newblock Numerical study of a nonlocal model for water-waves with variable
      depth.
    \newblock {\em Wave Motion}, 50:80--93, 2013.
    
    \bibitem{B}
    T.B. Benjamin.
    \newblock Instability of periodic wavetrains in nonlinear dispersive systems.
    \newblock {\em Proceedings of the Royal Society of London, Series A},
      299(1456):59--76, 1967.
    
    \bibitem{BenF}
    T.B. Benjamin and J.E. Feir.
    \newblock The disintegration of wave trains on deep water: {P}art {I}.
      {T}heory.
    \newblock {\em Journal of Fluid Mechanics}, 27:417--430, 1967.
    
    \bibitem{Binswanger}
    A.L. Binswanger, M.A. Hoefer, B.~Ilan, and P.~Sprenger.
    \newblock Whitham modulation theory for generalized {W}hitham equations and a
      general criterion for modulational instability.
    \newblock {\em Studies in Applied Mathematics}, 2021.
    
    \bibitem{BD}
    N.~Bottman and B.~Deconinck.
    \newblock {KdV} cnoidal waves are linearly stable.
    \newblock {\em Discrete and Continuous Dynamical Systems A}, 25:1163--1180,
      2009.
    
    \bibitem{WhithamComp}
    J.D. Carter.
    \newblock Bidirectional {W}hitham equations as models of waves on shallow
      water.
    \newblock {\em Wave Motion}, 82:51--61, 2018.
    
    \bibitem{STWhitham}
    J.D. Carter and M.~Rozman.
    \newblock Stability of periodic, traveling-wave solutions to the
      capillary-{W}hitham equation.
    \newblock {\em Fluids}, 4:58, 2019.
    
    \bibitem{DK}
    B.~Deconinck and J.N. Kutz.
    \newblock Computing spectra of linear operators using {H}ill's method.
    \newblock {\em Journal of Computational Physics}, 219:296--321, 2006.
    
    \bibitem{BernardOlga}
    B.~Deconinck and O.~Trichtchenko.
    \newblock High-frequency instabilities of small-amplitude {H}amiltonian {PDE}s.
    \newblock {\em Discrete and Continuous Dynamical Systems}, 37(3):1323--1358,
      2015.
    
    \bibitem{Dinvay}
    E.~Dinvay, D.~Moldabayev, D.~Dutykh, and H.~Kalisch.
    \newblock The {W}hitham equation with surface tension.
    \newblock {\em Nonlinear Dynamics}, 88:1125--1138, 2017.
    
    \bibitem{EK}
    M.~Ehrnstr\"om and H.~Kalisch.
    \newblock Traveling waves for the {W}hitham equation.
    \newblock {\em Differential and Integral Equations}, 22:1193--1210, 2009.
    
    \bibitem{WhithamCusp}
    M.~Ehrnstr\"om and E.~Wahl\'en.
    \newblock On {W}hitham's conjecture of a highest cusped wave for a nonlocal
      dispersive equation.
    \newblock {\em Annales de l'Institut Henri Poincare. Analyse non lin\'ear},
      36:769--799, 2019.
    
    \bibitem{h}
    J.~Hammack.
    \newblock A note on tsunamis: their generation and propagation in an ocean of
      uniform depth.
    \newblock {\em Journal of Fluid Mechanics}, 60:769--799, 1973.
    
    \bibitem{hs}
    J.L. Hammack and H.~Segur.
    \newblock The {K}orteweg-de {V}ries equation and water waves. {P}art 2.
      {C}omparison with experiments.
    \newblock {\em Journal of Fluid Mechanics}, 65:289--314, 1974.
    
    \bibitem{HurJohnson2015}
    V.M. Hur and M.A. Johnson.
    \newblock Modulational instability in the {W}hitham equation for water waves.
    \newblock {\em Studies in Applied Mathematics}, 134(1):120--143, 2015.
    
    \bibitem{Hur2015}
    V.M. Hur and M.A. Johnson.
    \newblock Modulational instability in the {W}hitham equation with surface
      tension and vorticity.
    \newblock {\em Nonlinear Analysis}, 129:104--118, 2015.
    
    \bibitem{HP}
    V.M. Hur and A.K. Pandey.
    \newblock Modulational instability in a full-dispersion shallow water model.
    \newblock {\em Studies in Applied Mathematics}, 142:3--47, 2019.
    
    \bibitem{johnson}
    R.S. Johnson.
    \newblock {\em A Modern Introduction to the Mathematical Theory of Water
      Waves}.
    \newblock Cambridge University Press, 2001.
    
    \bibitem{Kharif2018}
    C.~Kharif and M.~Abid.
    \newblock Nonlinear water waves in shallow water in the presence of constant
      vorticity: A {W}hitham approach.
    \newblock {\em European Journal of Mechanics B/ Fluids}, 72:12--22, 2018.
    
    \bibitem{vorW}
    C.~Kharif, M.~Abid, J.D. Carter, and H.~Kalisch.
    \newblock Stability of periodic progressive gravity wave solutions of the
      {W}hitham equation in the presence of vorticity.
    \newblock {\em Physics Letters A}, 384:126060, 2020.
    
    \bibitem{Kharif2017}
    C.~Kharif, M.~Abid, and J.~Touboul.
    \newblock Rogue waves in shallow water in the presence of a vertically sheared
      current.
    \newblock {\em Journal of Ocean Engineering and Marine Energy}, 3:301--308,
      2017.
    
    \bibitem{kdv}
    D.J. Korteweg and D.~de~Vries.
    \newblock On the change of form of long waves advancing in a rectangular canal,
      and on a new type of long stationary wave.
    \newblock {\em Philosophical Magazine}, 39:422--443, 1895.
    
    \bibitem{LannesBook}
    D.~Lannes.
    \newblock {\em The Water Waves Problem: Mathematical Analysis and Asymptotics}.
    \newblock American Mathematical Society, 2013.
    
    \bibitem{McLean}
    J.W. McLean.
    \newblock Instabilities of finite-amplitude gravity waves on water of finite
      depth.
    \newblock {\em Journal of Fluid Mechanics}, 114:331--341, 1982.
    
    \bibitem{miles}
    J.W. Miles.
    \newblock The {K}orteweg-de {V}ries equation, a historical essay.
    \newblock {\em Journal of Fluid Mechanics}, 106:131--147, 1981.
    
    \bibitem{Moldabayev}
    D.~Moldabayev, H.~Kalisch, and D.~Dutykh.
    \newblock The {W}hitham equation as a model for surface water waves.
    \newblock {\em Physica D}, 309:99--107, 2015.
    
    \bibitem{russell}
    J.S. Russell.
    \newblock Report on waves.
    \newblock {\em Report of the fourteenth meeting of the British Association for
      the Advancement of Science, York}, pages 311--390, 1844.
    
    \bibitem{Sanford2014}
    N.~Sanford, K.~Kodama, J.D. Carter, and H.~Kalisch.
    \newblock Stability of traveling wave solutions to the {W}hitham equation.
    \newblock {\em Physics Letters A}, 378:2100--2107, 2014.
    
    \bibitem{Tanaka}
    M.~Tanaka, J.W. Dold, M.~Lewy, and D.H. Peregrine.
    \newblock Instability and breaking of solitary waves.
    \newblock {\em Journal of Fluid Mechanics}, 185:235--248, 1987.
    
    \bibitem{Trillo}
    S.~Trillo, M.~Klein, G.F. Clauss, and M.~Onorato.
    \newblock Observation of dispersive shock waves developing from initial
      depressions in shallow water.
    \newblock {\em Physica D}, 333:276--284, 2016.
    
    \bibitem{Whitham}
    G.B. Whitham.
    \newblock Variational methods and applications to water waves.
    \newblock {\em Proceedings of the Royal Society of London, A}, 299:6--25, 1967.
    
    \bibitem{Whithambook}
    G.B Whitham.
    \newblock {\em Linear and Nonlinear Waves}.
    \newblock John Wiley \& Sons, Inc., New York, 1974.
    
    \bibitem{zabusky}
    N.J. Zabusky and C.J. Galvin.
    \newblock Shallow-water waves, the {K}orteweg-de {V}ries equation and solitons.
    \newblock {\em Journal of Fluid Mechanics}, 47:811--824, 1971.
    
    \end{thebibliography}
\end{document}